\documentclass[prl,twocolumn,preprintnumbers,amsmath,amssymb]{revtex4-1}

\usepackage{upgreek,graphicx}

\begin{document}

\title{Transport in quenched disorder: light diffusion in strongly heterogeneous turbid media}

\author{Tomas Svensson$^1$}
\email[Email: ]{svensson@lens.unifi.it}
\author{Kevin Vynck$^{1,2}$}
\author{Erik Adolfsson$^3$}
\author{Andrea Farina$^4$}
\author{Antonio Pifferi$^{4,5}$}
\author{Diederik S. Wiersma$^{1,6}$}

\affiliation{$^1$European Laboratory for Non-linear Spectroscopy (LENS), University of Florence, Via Nello Carrara 1, 50019 Sesto Fiorentino, Italy}
\affiliation{$^2$Institut Langevin, ESPCI ParisTech, CNRS, 1 rue Jussieu, 75238 Paris Cedex 05, France}
\affiliation{$^3$Ceramic Materials, SWEREA IVF, P.O. Box 104,SE- 431 22 M\"{o}lndal, Sweden}
\affiliation{$^4$\mbox{Istituto di Fotonica e Nanotecnologie (IFN-CNR), Piazza Leonardo da Vinci 32, Milan 20133 Italy}}
\affiliation{$^5$Dipartimento di Fisica, Politecnico di Milano, Piazza Leonardo da Vinci 32, Milan 20133, Italy}
\affiliation{$^6$Istituto Nazionale di Ottica (CNR-INO), Largo Fermi 6, 50125 Firenze, Italy}

\date{\today}

\begin{abstract}
We present a theoretical and experimental study of light transport in disordered media with strongly heterogeneous distribution of scatterers formed via non-scattering regions. Step correlations induced by quenched disorder are found to prevent diffusivity from diverging with increasing heterogeneity scale, contrary to expectations from annealed models. Spectral diffusivity is measured for a porous ceramic where nanopores act as scatterers and macropores render their distribution heterogeneous. Results agree well with Monte Carlo simulations and a proposed analytical model.
\end{abstract}


\maketitle

Series of incremental random changes govern the evolution of countless systems around us, from the movement of particles and molecules \cite{Brown1828_PhilMag,Einstein1905b_AnnPhys,Berg1993_Book,Cussler2009_Book} and wave propagation in disordered media \cite{Sheng2006_Book,Akkermans2007_Book}, to the foraging of animals \cite{Viswanathan2011_Book} and spread of disease \cite{Brockmann2006_Nature}. By virtue of the central limit theorem, macroscopic evolution of such random walk processes can often be explained in terms of classical diffusion. Although therefore ubiquitous, diffusion is in each particular case determined by unique microscopic mechanisms. These mechanisms are often complex and understanding the onset and speed of diffusion is generally a challenge \cite{Bouchaud1990_PhysRep,Elaloufi2004_JOSAA,Lukic2005_PRL,Huang2011_NatPhys,Novikov2011_NatPhys,Wang2012_NatMater}. The matter is particularly relevant to research in optics of disordered media, including the study of radiative transfer through planetary atmospheres \cite{Chandrasekhar1960_Book,Hansen1974_SpaceSciRev,Davis2010_RepProgPhys}, optical imaging and spectroscopy in biomedical \cite{Welch2010_Book} and material science \cite{Berne2000_Book} and, more recently, anomalous diffusion in engineered disordered materials \cite{Barthelemy2008_Nature,Barthelemy2010_PRE,Vynck2012_Chapter, Buonsante2011_PRE, Groth2012_PRE}. 

Multiple scattering of light is typically viewed as a Poissonian random walk of independent and exponentially-distributed steps. This viewpoint inherently assumes a uniform random distribution of scatterers throughout the medium and results in a well-known expression for the diffusion constant \cite{Rossum1999_RevModPhys}, $D=v \ell_t/3$, where $v$ is the average transport velocity for light in the medium and $\ell_t$ the transport mean free path. This diffusivity relation, however, breaks down in systems with an \textit{heterogeneous} distribution of scatterers, such as clouds, biological tissues, porous materials and foams. The reason is two-fold. First, the presence of non-scattering regions in the scattering medium leads to a broader (non-exponential) distribution of step lengths, which, in turn, induces an increase of the diffusivity \cite{Svensson2013_PRE}. Second, the quenched (i.e. spatially frozen) heterogeneity induces step correlations that tend to counteract the  increase in diffusivity caused by long steps \cite{Barthelemy2010_PRE,Svensson2013_PRE}. Due to the complexity of these aspects, understanding of light transport in systems with heterogeneous distribution of scatterers remains rather limited. Important insight has, nonetheless, been reached with the development of generalized transport equations and homogenization theory~\cite{Davis1997_FractalFrontiers,Kostinski2001_JOSAA,Bal2002_JCompPhys, Davis2004_JQSRT, Scholl2006_JGeophysRes, Frank2010_KinetRelatModels, Davis2011_JQSRT, Larsen2011_JQSRT, Lovejoy2009_PhysA} as well as probabilistic analysis of random walks~\cite{Behrens1949_ProcPhysSocA,Lieberoth1980_NuclSciEng, Svensson2013_PRE}. When it comes to quenched disorder, most works are theoretical and fall within the context of anomalous diffusion \cite{Fogedby1994_PRL, Schulz2002_PhysLettA, Burioni2010_PRE, Barthelemy2010_PRE, Buonsante2011_PRE, Groth2012_PRE}. Experimental and theoretical investigations of how regular diffusion is affected by quenched (frozen) disorder and the accompanying step correlations are, on the other hand, largely missing.

\begin{figure}[h]
  \includegraphics[]{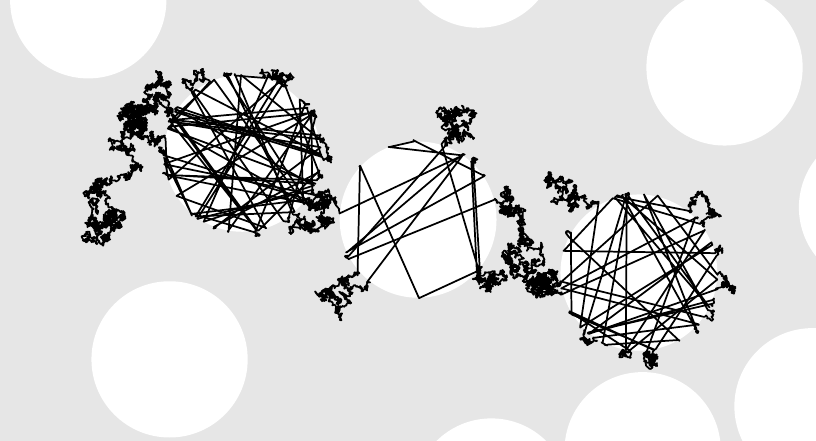}\\
  \caption{Monte Carlo simulation of light transport in a two-dimensional holey system. Non-scattering regions embedded into a turbid medium form a type of quenched disorder that, depending on the properties of the turbid medium, can induce strong step correlations. Here, in similarity to our 3D experiments on porous ceramics, the diameter of the non-scattering regions is 180 times larger than the transport mean free path of the turbid medium, resulting in strong step correlations.}\label{FIG1_HoleyRW}
\end{figure}

In this Letter, we theoretically and experimentally investigate light diffusion in heterogeneous systems constituted by turbid media with embedded non-scattering regions (holes). Special focus is on how quenched disorder influence transport. The transport process, which can be referred to as a \textit{holey random walk} \cite{Svensson2013_PRE}, is illustrated in Fig. \ref{FIG1_HoleyRW}. 
An analytical model for the diffusivity in such media is developed and compared to direct Monte Carlo simulations (using sphere packings to define hole arrangement). An important finding is that step correlations prevent the diffusivity to diverge with increasing heterogeneity size, contrary to expectations from annealed models. Experiments are conducted on a custom made ceramic with a bimodal pore size distribution: a nanoporous, strongly scattering base with embedded macropores acting as holes. Time-resolved measurements of light transmission allows assessment of diffusivity in the 600-900 nm spectral range, and results are found to be in agreement with theory.

\begin{figure}[h]
 \includegraphics[]{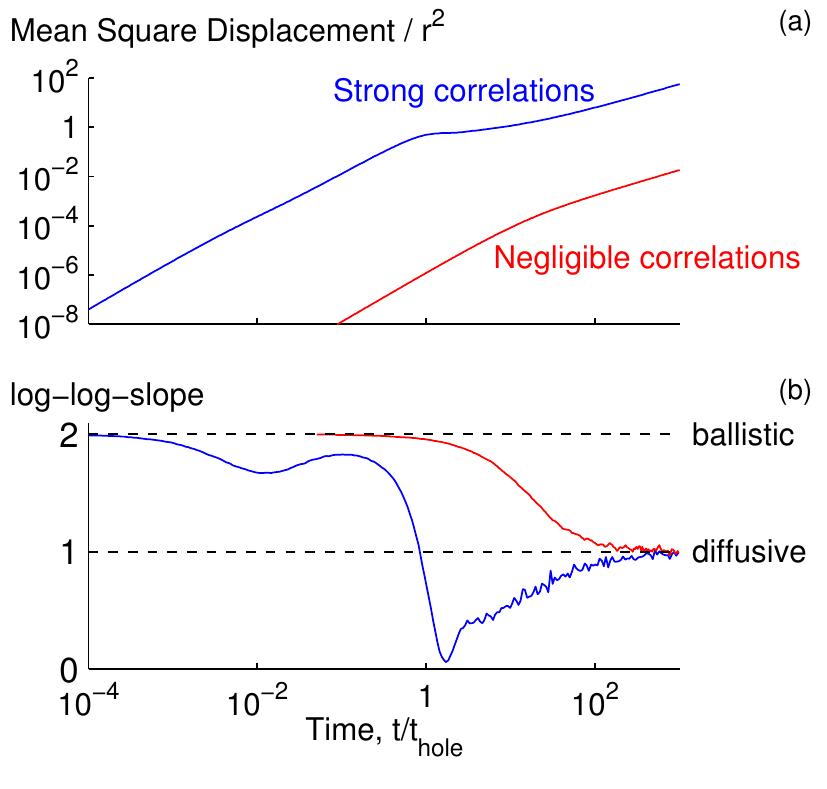}\\
  \caption{Dynamics of transport in holey systems illustrated via MSD evolution (time relative to the hole crossing time $t_\textrm{hole}=2r/v$). Panel (a) shows the MSD and panel (b) the log-log-derivative (which equals 2 and 1 for ballistic and diffusive transport, respectively). When $\ell_t \gg r$, step correlations are negligible and MSD evolution changes smoothly from ballistic to diffusive. In contrast, when $\ell_t\ll r$, steps are strongly correlated and diffusive evolution is reached via transient subdiffusive dynamics (the log-log-slope can even be below zero).}
  \label{FIG2_MSD}
\end{figure}

The system under consideration is a turbid medium characterized by a transport mean free path $\ell_t$ and containing non-scattering spherical holes of radius $r$ at filling fraction $\phi$. The holes do not overlap and may be arranged randomly or periodically.  In the limit of $\ell_t \gg  r$, step correlations become negligible and an (approximate) analytical expression for the resulting diffusion constant can be calculated from the ratio of the mean squared step ($E[S^2]$) to the mean step ($E[S]$) \cite{Svensson2013_PRE}. In general, however, the resulting diffusivity is not known. As $\ell_t$ approaches and becomes smaller than $r$ (cf. Fig. \ref{FIG1_HoleyRW}), step correlations get increasingly important and the diffusion constant becomes difficult to assess. To gain insight into this matter, we performed a series of Monte Carlo simulations (MC) of random walks in three-dimensional systems containing randomly or periodically arranged holes with $\phi=0.3$. The diffusion constant can be estimated by looking at the evolution of the mean square displacement (MSD) at long times. In the diffusive limit, the MSD is linear with time, $\textrm{MSD}=6Dt$ (in 3D). Figure~\ref{FIG2_MSD} shows the time-dependent mean square displacement (MSD) for the cases of both strong and negligible step correlations ($\ell_t \approx 0.005r$ and $\ell_t \approx 10r$, respectively). In the latter case, the transition from ballistic to diffusive transport is smooth and similar to that observed in most random walks with independent increments. In great contrast, the presence of strong step correlations results in a MSD evolution that goes from ballistic to diffusive via a \textit{transient subdiffusive} behavior (cf. \cite{Barthelemy2010_PRE}). This behavior is caused by the fact that long steps through holes often are counteracted by steps back through the same hole and constitutes a fingerprint of step correlations in disordered media (cf. trajectories in Fig. \ref{FIG1_HoleyRW}).

\begin{figure}[h]
  \includegraphics[]{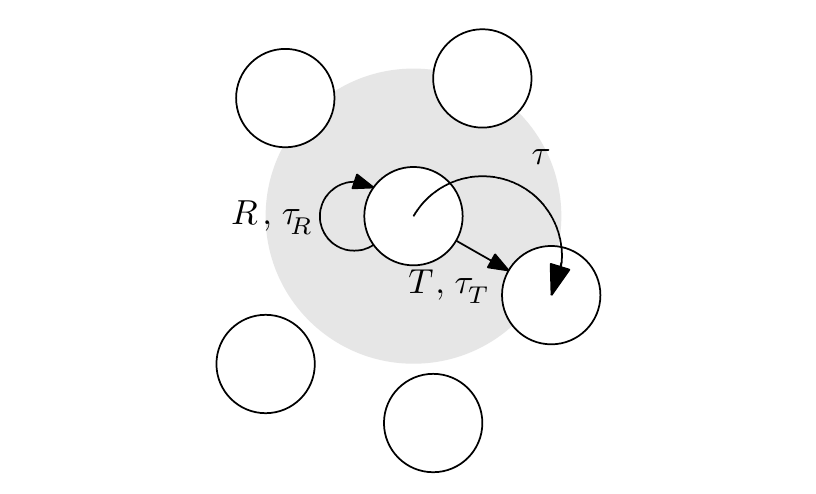}\\
  \caption{Lattice model for transport in holey media. Transport can be viewed as hops between holes, i.e. a random walk on a lattice. The hole-hole transfer time $\tau$ can be expressed as a function of the return probability $R=1-T$, the average return time $\tau_R$ and the average transmission time $\tau_T$. If the return probability is large, we propose that these unknown parameters can be calculated from a diffusion model for a spherical shell (schematically added as a shaded region).}
  \label{FIG3_LatticeModel}
\end{figure}

Let us now focus on the fact that strong step correlations goes hand in hand with random walkers being likely to return to the same hole several times before reaching a new one. We propose that transport in such systems can be viewed as a random walk on a lattice (similar to the approach used to understand deterministic diffusion in a periodic Lorentz gas~\cite{Machta1983_PRL,Klages2000_JStatPhys}), where the holes correspond to the lattice sites, with $a$ the lattice constant and $\tau$ the hole-hole transfer time. The diffusivity is then simply given by $D=a^2/6\tau$~\cite{Ben-Avraham2000}. The difficult part consists in determining $a$ and $\tau$. We propose that (i) the lattice constant $a$ can be approximated by the lattice constant of a periodic face-centered cubic lattice, $a = \sqrt[3]{\frac{8\pi r^3}{3\sqrt{2}\phi}}$, and (ii), in the limit when the distance between voids is much larger than $\ell_t$, the transfer time $\tau$ can be obtained from classical diffusion in a spherical shell with diffusivity $D_\textrm{shell}=v\ell_t/3$. The analytical model as a whole is illustrated in Fig.~\ref{FIG3_LatticeModel}. As we shall see and discuss below, although the model relies on several important approximations, it grasps the essential physics of the process and allows quantitative comparison with MC and experiments.

The transfer time $\tau$ is a sum of (i) the time spent on returning to the hole $\tau_R$, (ii) the time to cross the hole when first reaching it and after returning to it, and (iii) the time $\tau_T$ spent on reaching a new hole once the return series is broken. The number of returns before transfer to a new hole follows a geometric distribution, on average being $R/(1-R)=R/T$, and the number of crossings equals $R/T+1=1/T$. We therefore reach
\begin{align}\label{eq:transfer_time}
   \tau = \frac{R}{T}\times\tau_R + \frac{1}{T}\times\frac{E[\zeta]}{v_h}  + \tau_T,
\end{align}
where $E[\zeta]$ is the average chord length ($=4r/3$ for spheres, assuming isotropic flux) and $v_h$ the light velocity in the hole. Analytical expressions for the unknown parameters in Eq.~\ref{eq:transfer_time} were derived by solving the diffusion equation in spherical shells (derivation given in Supplemental Material) and are summmarized here. $\tau_R$ and $\tau_T$ can be retrieved from Eq.~\ref{eq:tau_RT} via $\tau_R=\tilde{\tau}(r)$ and $\tau_T=\tilde{\tau}(r+L_p)$, and $R$ and $T$ from Eq.~\ref{eq:RT} via $R=| F(r) |$ and $T=| F(r+L_p) |$. There, $\textrm{Li}_s(z)=\sum_{k=1}^\infty\frac{z^k}{k^s}$ is the Jonqui\`ere's function (the polylogarithm) \cite{Jonquiere1889_BSocMathFr}, $r_0=r+\ell_t$, $L_p$ is the physical thickness of the spherical shell, and $L=L_p+2r_e$ the extrapolated thickness ($r_e$ being the extrapolation length that can be calculated as done for light diffusion in, e.g., planar slabs~\cite{Contini1997_ApplOpt}). The remaining step is to set the shell thickness $L_p$ adequately. The actual diffusive process between spherical holes being particularly complex, the derivation of an exact value for $L_p$ that would take into account all characteristics of the medium (hole arrangement, filling fraction, etc) seems out-of-reach. Nevertheless, setting $L_p$ between $a-2r$ (smallest gap) and $a-r$ (distance from hole edge to center of next hole) appears reasonable and works well for the structures considered here. Above all, the exact value of $L_p$ has little effect on the relation between diffusivity and hole size reported below, which is a main message of the paper.
\begin{widetext}
\begin{align}
   \tilde{\tau}(\tilde{r}) =& \frac{L^2}{D_\textrm{shell}\pi^2} \frac{ -\textrm{Li}_3\big(e^{-i\frac{\pi}{L}(\tilde{r}-r_0)}\big) +\textrm{Li}_3\big(e^{i\frac{\pi}{L}(\tilde{r}-r_0)}\big) +\textrm{Li}_3\big(e^{-i\frac{\pi}{L}(\tilde{r}+r_0-2(r-r_e))}\big) -\textrm{Li}_3\big(e^{i\frac{\pi}{L}(\tilde{r}+r_0-2(r-r_e))}\big)}{ -\textrm{Li}_1\big(e^{-i\frac{\pi}{L}(\tilde{r}-r_0)}\big) +\textrm{Li}_1\big(e^{i\frac{\pi}{L}(\tilde{r}-r_0)}\big) +\textrm{Li}_1\big(e^{-i\frac{\pi}{L}(\tilde{r}+r_0-2(r-r_e))}\big) -\textrm{Li}_1\big(e^{i\frac{\pi}{L}(\tilde{r}+r_0-2(r-r_e))}\big)} \label{eq:tau_RT} \\
   F(\tilde{r}) =& \frac{-i\pi \tilde{r}}{2\pi^2r_0} \bigg(  -\textrm{Li}_1\big(e^{-i\frac{\pi}{L}(\tilde{r}-r_0)}\big) +\textrm{Li}_1\big(e^{i\frac{\pi}{L}(\tilde{r}-r_0)}\big) +\textrm{Li}_1\big(e^{-i\frac{\pi}{L}(\tilde{r}+r_0-2(r-r_e))}\big) -\textrm{Li}_1\big(e^{i\frac{\pi}{L}(\tilde{r}+r_0-2(r-r_e))}\big) \bigg) + \nonumber \\
   & \frac{L}{2\pi^2r_0} \bigg( -\textrm{Li}_3\big(e^{-i\frac{\pi}{L}(\tilde{r}-r_0)}\big) +\textrm{Li}_3\big(e^{i\frac{\pi}{L}(\tilde{r}-r_0)}\big) +\textrm{Li}_3\big(e^{-i\frac{\pi}{L}(\tilde{r}+r_0-2(r-r_e))}\big) -\textrm{Li}_3\big(e^{i\frac{\pi}{L}(\tilde{r}+r_0-2(r-r_e))}\big) \bigg) \label{eq:RT}
\end{align}
\end{widetext}

\begin{figure}
  \includegraphics[]{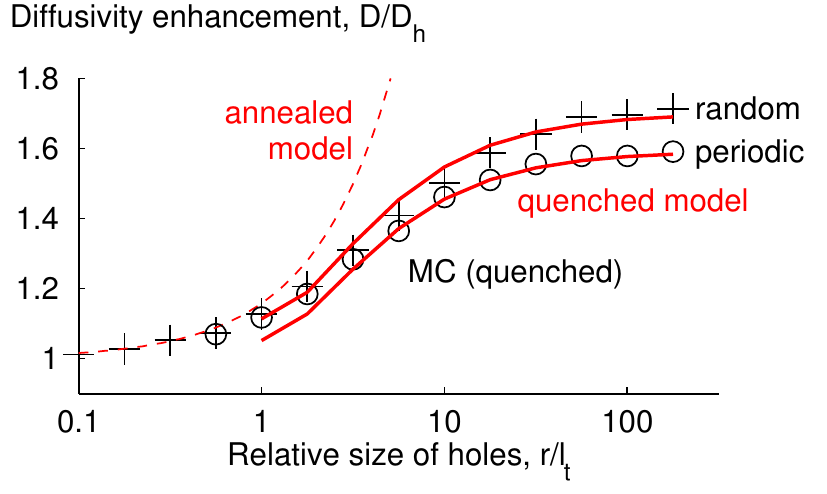}\\
  \caption{Importance of step correlations on transport. $E[S^2]$ diverges with increasing $r/\ell_t$, but step correlations prevent $D$ from diverging. Neglecting step correlations (annealed model in Ref.~\cite{Svensson2013_PRE}), one would obtain the growth in $D/D_h$ shown by the red dashed line. MC of quenched three-dimensional systems with periodically or randomly arranged holes (markers) show that diffusion is only up 1.6-1.7 times faster than if scatterers were homogeneously distributed. In the limit of significant step correlations, our analytical model (red solid lines) agrees quantitatively with MC, when $L_p$ is set to $a - 1.6r$ and $a - 1.65r$ for the periodic and random systems, respectively.}
  \label{FIG4_MC_IndexMatched}
\end{figure}

Figure~\ref{FIG4_MC_IndexMatched} shows how diffusivity is modified with increasing hole size (index-matched system, $\phi=0.3$ fixed), see note \footnote{Note that for a given hole filling fraction $\phi$, the diffusivity will depend essentially on the ratio $r/\ell_t$. This follows from a scaling consideration: when scaling the holey system and random walk with a factor $\gamma$, we have that $r \rightarrow \gamma r$ and $\ell_t \rightarrow \gamma \ell_t$ and $t \rightarrow \gamma t$. For diffusion constants, we therefore will have $D(\gamma r, \gamma \ell_t,\phi)=\gamma D(r,\ell_t, \phi)$.}. Diffusivity is reported with respect to the homogenized counterpart, i.e. the diffusion constant for a system with the same amount of scatterers but distributed homogeneously, $D_h = \frac{v\ell_t/3}{1-\phi}$~\cite{Svensson2013_PRE}. The outcome of MC and of the analytical model presented above are given by the markers and the solid lines respectively. Clearly, the model captures the essence of the transport process (for large $r/\ell_t$). For the periodic and random holey systems, quantitative agreement occurs when setting the shell thickness to $L_p = a - 1.6r$ and $L_p = a - 1.65r$, respectively (i.e. midway between the limits introduced above). The diffusivity enhancement evaluated from the annealed model (Ref.~\cite{Svensson2013_PRE}) is also shown for comparison (dashed line). The "annealed" diffusion constant diverges due to a diverging mean squared step length $E[S^2]$ (the mean step length, $E[S]$ remaining constant), while the "quenched" diffusion constants saturate at a value of about 1.6-1.7 times that of the homogenized system. This leads us to a very important conclusion: In quenched disordered systems, step correlations become so strong with increasing heterogeneity size that they completely counteract the increase of $E[S^2]$, thereby preventing the diffusivity to diverge. It is interesting to realize that in a one-dimensional system, $D=D_h$ regardless of how scatterers are distributed. The average number of scatterers that needs to be passed over macroscopic distances remains the same. Clearly, the situation is very different in three-dimensional systems.

We now test this prediction experimentally. We have manufactured porous ceramics with a bimodal pore size distribution by sintering a mixture of a zirconia nano particles, latex nanoparticles and 180 $\upmu$m diameter PMMA microspheres. Manufacturing details are given in Supplemental Material. Briefly, latex nanoparticles and microspheres are burned out during sintering, leaving a nanoporous ceramic with embedded macropores occupying around 30\% of the total volume (the porosity of the nanoporous part being 46\%). The latex particles are used as spacers, increasing the final distance between the different zirconia particles and, as a result, also increasing light scattering (see Ref.~\cite{Svensson2013_APB}). The nanoporous part acts as the turbid medium, and by manufacturing a reference sample without any macropores, its $\ell_t$ can be measured. The optical properties of these materials are studied via optical time-of-flight spectroscopy~\cite{Patterson1989_ApplOpt, Svensson2011_PRL}. Short picosecond pulses are injected into the sample and the diffuse transmission is resolved in time using time-correlated single photon counting. The system used is described in detail in Refs.~\cite{Bassi2007_OptExpress,Bargigia2012_ApplSpectrosc} and allows coverage of the 600-900 nm spectral range. The diffusion constant (and absorption coefficient) can be determined from the temporal shape of the transmitted pulse. Using a reasonable estimate of the effective refractive index $n_\textrm{eff}$ (here, we use the averaged permittivity), $\ell_t$ of the nanoporous part can be estimated from the measured $D=D_\textrm{ref}$ of the reference sample (via $D_\textrm{ref}=v\ell_t/3$ with $v=c_0/n_\textrm{eff}$). Accordingly, the turbid material between holes, is found to exhibit $\ell_t$ ranging from 0.7$~\upmu$m at 600 nm to 2.2$~\upmu$m at 900 nm and low absorption (0.026 cm$^{-1}$ at 660 nm down to 0.005 cm$^{-1}$ at 900 nm). The small size of the nanopores makes scattering strongly wavelength-dependent, and the transport scattering coefficient $\mu_s'$ decays as $\sim\lambda^{-2.7}$ (in good agreement with previous reports on scattering of nanoporous ceramics~\cite{Svensson2010_APL, Svensson2011_PRL}).

\begin{figure}
  \includegraphics[]{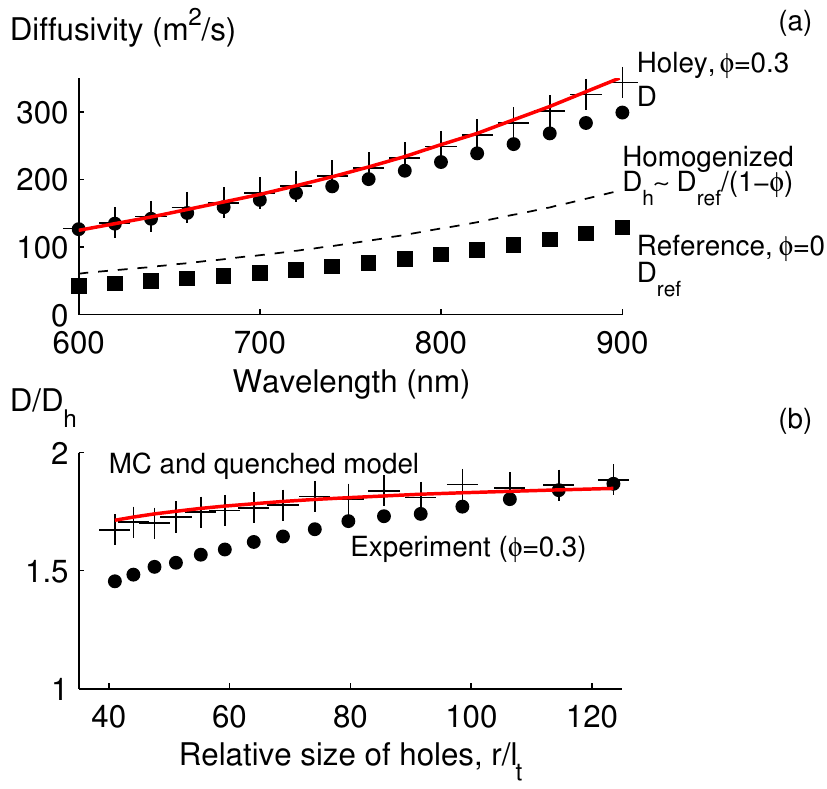}\\
  \caption{Experimental results and comparison with theory and simulations. Panel (a) shows the measured diffusivity of the holey ceramic (nanoporous material with macropores, $\phi=0.3$, $r=90~\upmu$m; dots) and a purely nanoporous reference (squares). Diffusivity predicted by MC ($+$) and our lattice model (red solid line) is also shown, as well as the homogenized counterpart of the holey system (dashed line). Clearly, homogenization is not an appropriate approximation of macroscopic transport in holey systems, while a good agreement is found with our theory and simulations on quenched holey system. Along the lines of Fig.~\ref{FIG4_MC_IndexMatched}, panel (b) shows the diffusivity enhancement $D/D_h$ for the holey system.}
  \label{FIG5_Experimental}
\end{figure}

Figure~\ref{FIG5_Experimental}a shows the diffusivity spectra for the holey (bimodal) system and for the nanoporous reference, and compares it to MC and the proposed analytical model. The refractive index mismatch between holes and nanoporous media is taken into account. The discrepancy between experiments and theory may be due to differences in $\ell_t$ between the reference and the holey system (scattering is very sensitive to the microstructure and differences related to the inclusion and burnout of microspheres cannot be ruled out) or to the inaccuracy in extracting $\ell_t$ from $D_\textrm{ref}$ experimentally. Clearly, the holey system exhibits significantly faster diffusion than its homogenized counterpart (here $D_h=\frac{v\ell_t}{3(1-\phi)}\times \frac{n_{\phi=0}}{n_{\phi=0.3}}$, where $n$ refers to the refractive index calculated from the averaged permittivity, in an attempt to take into account the change in average refractive index that follows from removal of solid material). At the same time, due to step correlations, diffusion is slower than what could be expected from the step length distribution. This is seen in Fig.~\ref{FIG5_Experimental}b, keeping in mind that if steps were uncorrelated, $D/D_h$ would diverge quickly (as in Fig.~\ref{FIG4_MC_IndexMatched}). The retrieved diffusivity enhancements are between 1.5 and 1.8, in agreement with our previous analysis. 

Let us also note that the non-linear relation between $D$ and $D_h$ has an important consequence: the shape of diffusivity spectra is not only dependent on the microstructure (e.g. scatterer or pore size). This has bearing on the interpretation of diffuse spectra in general. It is, for example, common to convert diffusivity spectra $D(\lambda)$ into a transport scattering spectra $\mu_s'(\lambda)$ and use that to assess particle size \cite{Mourant1997_ApplOpt}. This procedure relies on a linear relation between $D$ and $\ell_t=1/\mu_s'$, valid for homogeneously turbid media but, as shown here, not valid for holey media. While the nanoporous reference media has a diffusivity spectra that corresponds to a $\lambda^{-2.7}$ decay of $\mu_s'$, the holey system \--- when analyzed in the same way \--- exhibit a $\lambda^{-2.1}$ decay. This change in apparent scattering spectra is induced by heterogeneity, not by an increase of pore size.

To conclude, this work provides an initial understanding and a first experimental investigation of the importance of step correlations due to quenched disorder on diffusion in strongly heterogeneous media. We have found that step correlations can counteract completely the effect of a broad step length distribution, preventing diffusivity to diverge with increasing scale of heterogeneities. This effect was predicted theoretically, and confirmed experimentally and numerically. An important implication, relevant to analytical spectroscopy in general, is that strong heterogeneity complicates assessment of microstructural characteristics based on diffusivity spectra. In addition, besides the fundamental questions brought up in this study, we expect future challenges as the topic of light diffusion in strongly heterogeneous turbid media is generalized to include, for instance, non-spherical heterogeneities and/or anisotropic holey random walks (with relevance to, e.g., anisotropic diffusion in compressed porous matter \cite{Alerstam2012_PRE}). 

\begin{acknowledgments}
T.S. acknowledges funding from the Swedish Research Council (postdoctoral fellowship Grant No. 2010-887). K.V. acknowledges support by LABEX WIFI (Laboratory of Excellence within the French Program "Investments for the Future'') under references ANR-10-LABX-24 and ANR-10-IDEX-0001-02 PSL$^\star$. The reported research has also received funding from the European Research Council under the European UnionÕs Seventh Framework Programme (FP7/2007-2013) ERC grant agreement number 291349, and from LASERLAB-EUROPE (grant agreement number 284464). The authors are also grateful to Erik Alerstam, Matteo Burresi and Romolo Savo for long-time collaboration and fruitful discussions.
\end{acknowledgments}

\newpage

\onecolumngrid

\section{SUPPLEMENTAL MATERIAL}

\section{I. Diffusion through spherical shells}

Here, we derive an analytical expression for the energy density in diffusive spherical shells by solving the diffusion equation in spherical coordinates with Dirichlet boundary conditions and a Dirac delta impulse as an initial condition. Based on this, we derive expressions for time-resolved reflection and transmission. The theoretical results are compared with direct Monte Carlo simulations of transport in spherical shells. It should be noted that the notation used here differs from the notation used in the main article. In the main article $r$ refers to the radius of holes while it here refers to the spherical coordinate.

\subsection{The diffusion equation}

Conside a spherical shell of inner radius $r_{in}$ and outer radius $r_{out}$ containing a diffusive medium with diffusion constant $D$. A spherical plane source is placed at $r=r_0$ in order to have rotational invariance.
\begin{figure}[h]
\begin{center}
	\includegraphics[width=0.45\textwidth]{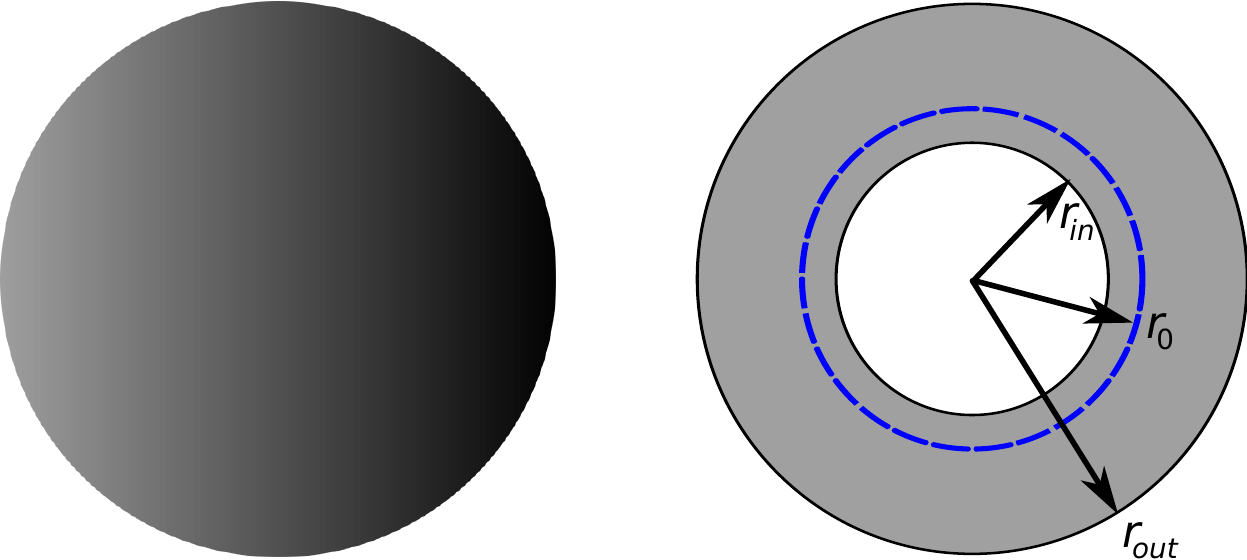}
\end{center}	
\caption{Cut of a spherical shell of diffusive medium with absorbing inner ($r=r_{in}$) and outer ($r=r_{out}$) boundaries and a spherical plane source ($r=r_{0}$).
\label{system}}
\end{figure}

The diffusion equation in its general form is written as
\begin{equation}
\frac{\partial u(\mathbf{r},t)}{\partial t} = D \nabla^2 u(\mathbf{r},t)
\end{equation}
We impose the Dirichlet boundary conditions $u(r=r_{in},t)=0$ and $u(r=r_{out},t)=0$, where $r=|\mathbf{r}|$ and take the initial condition $u(r,0)=\delta(r-r_0)$. 

In spherical coordinates, the Laplacian operator can be written as
\begin{equation}
\nabla^2 u=\frac{\partial^2 u}{\partial r^2} + \frac{2}{r} \frac{\partial u}{\partial r} + \frac{1}{r^2} \frac{\partial^2 u}{\partial \varphi^2} + \frac{\cos \varphi}{r^2 \sin \varphi} \frac{\partial u}{\partial \varphi} + \frac{1}{r^2 \sin^2 \varphi} \frac{\partial^2 u}{\partial \theta^2}
\end{equation}
and the energy density can be written by separating the variables as $u(\mathbf{r},t)=\Phi(r) X(\varphi,\theta) \Psi(t)$ to construct separable solutions of the diffusion equation. Since we assume rotational invariance, the angular terms disappear in the Laplacian and the energy density can then be written simply as
\begin{equation}
u(r,t)=\Phi(r) \Psi(t)
\end{equation}
and the diffusion equation in spherical coordinates, after division on each side by $\Phi(r) \Psi(t)$, reads
\begin{equation}
\frac{1}{\Psi(t)} \frac{\partial \Psi(t)}{\partial t} = \frac{D }{\Phi(r)} \frac{\partial^2 \Phi(r)}{\partial r^2} + \frac{2}{r} \frac{\partial \Phi(r)}{\partial r} = -\lambda
\end{equation}
where $\lambda$ is the separation constant. The solution of the temporal equation $\partial \Psi(t) / \partial t = -\lambda \Psi(t)$ simply gives $\Psi(t)=\exp \left( -\lambda t \right)$, such that $u(r,t)=\exp \left( -\lambda t \right) \Phi(r)$. At this point, the equation that remains to be solved is
\begin{equation}
\frac{\partial^2 \Phi(r)}{\partial r^2} + \frac{2}{r} \frac{\partial \Phi(r)}{\partial r} + \frac{\lambda}{D} \Phi(r)= 0
\end{equation}
We can use the known solution of the \textit{spherical Bessel differential equation}:
\begin{equation}
\frac{\partial^2 \Phi(r)}{\partial r^2} + \frac{2}{r} \frac{\partial \Phi(r)}{\partial r} + \left( \frac{\lambda}{D} - m(m+1)\right) \Phi(r)= 0
\end{equation}
with $m=0,1,2,3,...$ the eigenvalue order generally associated with the spherical harmonics. The solution of the equation is written in terms of \textit{spherical Bessel functions} as:
\begin{equation}
\Phi(r)=C_1 j_m \left[ \sqrt{\frac{\lambda}{D}} r \right]+C_2 y_m \left[ \sqrt{\frac{\lambda}{D}} r \right]
\end{equation}
where $C_1$ and $C_2$ are two constants depending on the boundary conditions, and $j_m$ and $y_m$ are the spherical Bessel functions of the first and second kind, respectively, of order $m$. Due to rotational invariance in our case, only the order $m=0$ remains. Interestingly, the spherical bessel functions of order 0 can be written as elementary functions: $j_0(x)=\sin(x)/x$ and $y_0(x)=-\cos(x)/x$. The solution of our equation can then be written as:
\begin{equation}
\Phi(r)=\sqrt{\frac{D}{\lambda}} \frac{1}{r} \left( C_1 \sin \left[ \sqrt{\frac{\lambda} {D}} r \right] - C_2 \cos \left[ \sqrt{\frac{\lambda} {D}} r \right] \right)
\end{equation}

\subsection{Boundary conditions}

We now apply the Dirichlet boundary conditions $\Phi(r=r_{in})=0$ and $\Phi(r=r_{out})=0$. By applying the former, we find that $C_2=C_1 \tan \left( \sqrt{\lambda/D} \: r_{in} \right)$, such that
\begin{equation}
\Phi(r)=C_1 \sqrt{\frac{D}{\lambda}} \frac{1}{r} \left( \sin \left[ \sqrt{\frac{\lambda}{D}} r \right] - \tan  \left[ \sqrt{\frac{\lambda}{D}} r_{in} \right] \cos \left[ \sqrt{\frac{\lambda}{D}} r \right] \right)
\end{equation}
and by applying the latter, we arrive at the following condition: $\tan[\sqrt{\lambda/D} \: r_{in}] = \tan[\sqrt{\lambda/D} \: r_{out}]$. Writing $r_{out}=r_{in}+L$ leads to the condition that $\sqrt{\lambda/D} L$ should be a multiple of $\pi$. Thus, the eigenvalues $\lambda$ should be:
\begin{equation}
\lambda_n=\frac{n^2 \pi^2}{L^2} D
\end{equation}
with $n=1,2,3,...$.

Since the eigenfunctions are complete on a bounded domain, the fundamental solution can be written as an eigenfunction series as $u(r,t)=\sum_{n=1}^\infty c_n \Psi_n(t) \Phi_n(r)$, in which the constant $C_1$ is contained within the coefficients $c_n$. We therefore have
\begin{equation}\label{eq:propag}
u(r,t)=\sum_{n=1}^\infty c_n \: \exp \left[ -\frac{n^2 \pi^2}{L^2} D t \right] \: \frac{L}{n \pi} \frac{1}{r} \left( \sin \left[ \frac{n\pi}{L} r \right] - \tan  \left[ \frac{n\pi}{L} r_{in} \right] \cos \left[ \frac{n\pi}{L} r \right] \right)
\end{equation}

\subsection{Initial conditions}

We now turn to the determination of the coefficients $c_n$, uniquely prescribed by the initial conditions at $t=0$ and are given by the usual orthogonality formula:
\begin{equation}
c_n=\frac{ \iiint_\Omega u(\mathbf{r},0) \Phi_n(\mathbf{r}) dV}{\iiint_\Omega \Phi_n^2(\mathbf{r}) dV}
\end{equation}
with $\Omega$ the integration volume. With $u(\mathbf{r},t_0)=\delta(|\mathbf{r}-\mathbf{r}_0|)$, we are left with two volume integrals to solve:
\begin{equation}
c_n=\frac{ \iiint_\Omega \delta(|\mathbf{r}-\mathbf{r}_0|) \Phi_n(\mathbf{r}) dV}{\iiint_\Omega \Phi_n^2(\mathbf{r}) dV}
\end{equation}

The volume element $dV$ in spherical coordinates and rotational invariance becomes $dV=4\pi r^2 dr$. The volume integral of the numerator is simply the function $4\pi \Phi(r_0) r_0^2$ and the denominator can be evaluated exactly. After some maths, we find:

\begin{equation}\label{eq:coeff}
c_n=\frac{2 n \pi r_0}{L^2}  \cos \left[ \frac{n\pi}{L} r_{in} \right] \sin \left[ \frac{n\pi}{L} (r_0-r_{in}) \right]
\end{equation}

\subsection{Propagator (energy density)}

Equations~(\ref{eq:propag},\ref{eq:coeff}) provide the complete solution for the propagator in diffusive spherical shells, which, all in all,  simplifies to:
\begin{equation}\label{eq:propag2}
u(r,t)=\sum_{n=1}^\infty \frac{2 r_0}{L r} \sin \left[ \frac{n\pi}{L} (r_0-r_{in}) \right] \sin \left[ \frac{n\pi}{L} (r-r_{in}) \right] \: \exp \left[ -\frac{n^2 \pi^2}{L^2} D t \right]
\end{equation}

For the application of Eq.~(\ref{eq:propag2}) to real samples, one should consider the fact that the intensity is not zero exactly at the boundary but at a distance $r_e=2/3 \ell (1+R_i)/(1-R_i)$, called the \textit{extrapolation length}, where $\ell$ is the mean free path in the diffusive medium and $R_i$ is the internal reflection coefficient at the boundary. See  \cite{Contini1997_ApplOpt} for a description on how to calculate the extrapolation length for a given mismatch in refractive index. It is possible to use Eq.~(\ref{eq:propag2}) directly by taking the thickness $L$ as the \textit{extrapolated} thickness $L=L_p+2r_e$, where $L_p$ is the \textit{physical} thickness. Similarly, the inner radius should be taken as an extrapolated inner radius $r_{in}=r_{in,p}-r_e$. Thus, a point source placed one mean free path apart from this boundary should be placed at $r_0=r_{in,p}+\ell=r_{in}+r_e+\ell$.

\begin{figure}[h]
\begin{center}
	\includegraphics[width=0.6\textwidth]{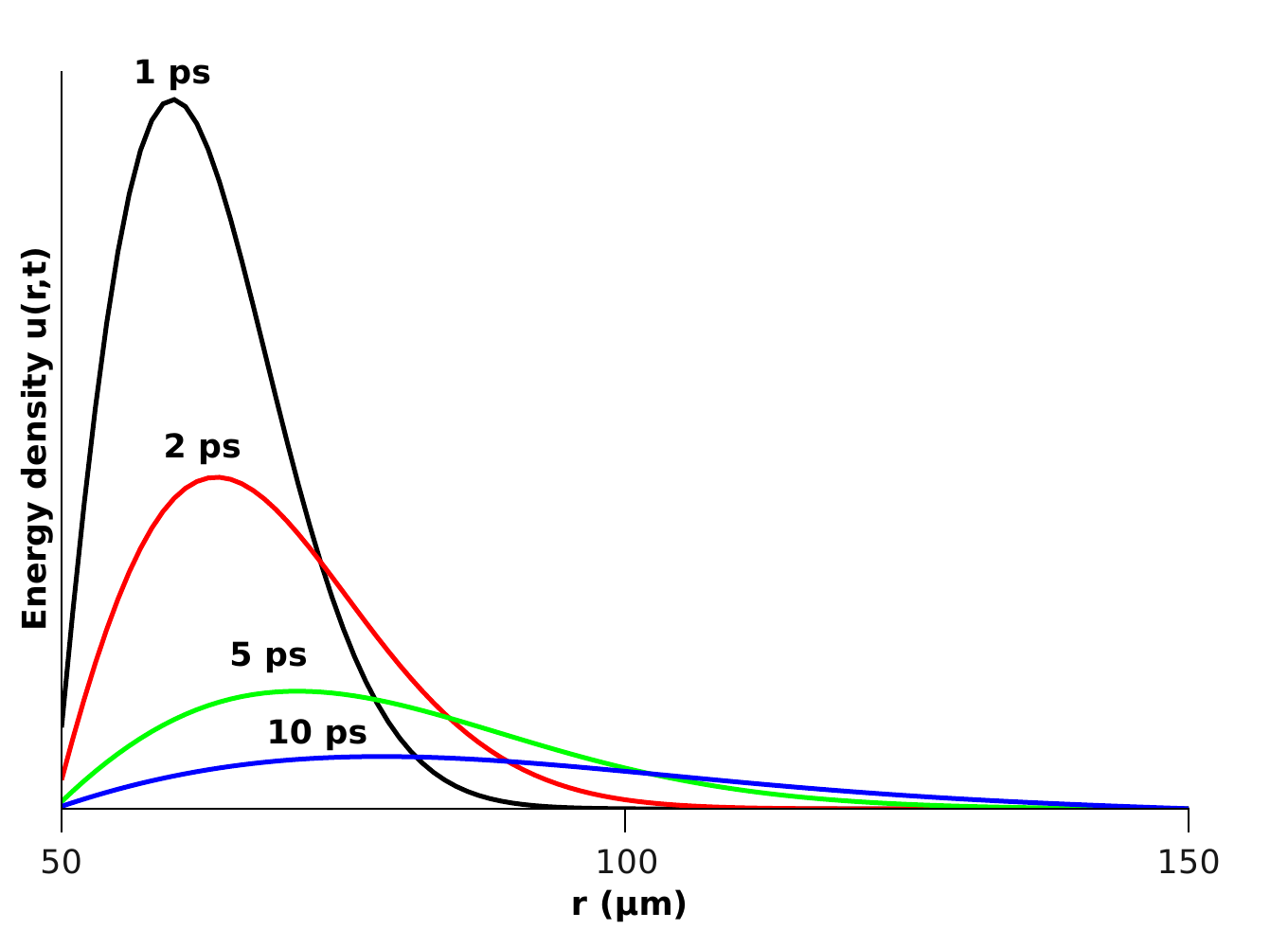}
\end{center}	
\caption{Temporal evolution of the energy density in a diffusive spherical shell.
\label{fig:energy_density}}
\end{figure}

The energy density in the spherical shell at different times is shown in Fig.~\ref{fig:energy_density} for a system with $r_{in,p}=50$ $\mu$m, $L_p=100$ $\mu$m, $\ell=1$ $\mu$m and $D=67$ $\mu$m$^2$/ps. The source is placed at $r_0=r_{in,p}+\ell=51$ $\mu$m.

\subsection{Energy flux}

To calculate the fluxes in transmission or reflection through the slab, it is necessary to apply the first Fick's law of diffusion:
\begin{equation}
\mathbf{J}=-D \nabla u(\mathbf{r},t)
\end{equation}

In spherical coordinates with rotational invariance, we have that $\nabla u(\mathbf{r},t)= \partial u(r,t) / \partial r \: \hat{\mathbf{r}}$.  Thus, the diffusion flux can easily be calculated by spatial derivation of $u(r,t)$ in Eq.~(\ref{eq:propag2}). We find:
\begin{equation}
J(r,t)=-D \sum_{n=1}^\infty \frac{2 r_0}{L r^2} \sin \left[ \frac{n\pi}{L} (r_0-r_{in}) \right] \left( \frac{n \pi }{L} r \cos \left[ \frac{n\pi}{L} (r-r_{in}) \right] - \sin \left[ \frac{n\pi}{L} (r-r_{in}) \right] \right) \: \exp \left[ -\frac{n^2 \pi^2}{L^2} D t \right]
\end{equation}

\subsection{Time-resolved transmission and reflection}

The time-resolved reflection and transmission are then found by integrating the fluxes over the surface of the corresponding boundaries and normalizing by the source $4\pi r_0^2$:
\begin{equation}
\begin{cases}
& R(t)=-\left( \frac{r_{in,p}}{r_0} \right)^2 J(r=r_{in,p}, t) \\
& T(t)=\left( \frac{r_{in,p} + L_p}{r_0} \right)^2 J(r=r_{in,p}+L_p, t)
\end{cases}
\end{equation}
The normalization $\int_0^\infty (R(t)+T(t)) dt =1$ can be verified.

\subsection{Mean first-passage time}

An important step is the calculation of the mean reflection and transmission (or first passage) time. This can be calculated as $\tau_R=\int_0^\infty t R(t) dt / \int_0^\infty R(t) dt$ and $\tau_T=\int_0^\infty t T(t) dt / \int_0^\infty T(t) dt$, respectively. An analytical expression can be found for this quantity by using the definition of the \textit{polylogarithm}, or Jonqui\`{e}re's function \cite{Jonquiere1889_BSocMathFr}, defined as $\text{Li}_s(z)=\sum_{k=1}^\infty \frac{z^k}{k^s}$, and Euler's formula on the sine and cosines. One finds (Eq. \ref{eq:tau_RT} in the main text):
\begin{equation}
\tau(r)=\frac{L^2}{D \pi^2} \frac{-\text{Li}_3 \left[ e^{-i \frac{\pi}{L}(r-r_0)} \right] + \text{Li}_3 \left[ e^{i \frac{\pi}{L}(r-r_0)} \right] + \text{Li}_3 \left[ e^{-i \frac{\pi}{L}(r+r_0-2r_{in})} \right] -\text{Li}_3 \left[ e^{i \frac{\pi}{L}(r+r_0-2r_{in})} \right]} {-\text{Li}_1 \left[ e^{-i \frac{\pi}{L}(r-r_0)} \right] + \text{Li}_1 \left[ e^{i \frac{\pi}{L}(r-r_0)} \right] + \text{Li}_1 \left[ e^{-i \frac{\pi}{L}(r+r_0-2r_{in})} \right] -\text{Li}_1 \left[ e^{i \frac{\pi}{L}(r+r_0-2r_{in})} \right]}
\end{equation}
In the practical example considered above, we find $\tau(r=r_{in,p})=\tau_R=0.834167$ ps and $\tau(r=r_{in,p}+L_p)=\tau_T=25.6608$ ps.

\subsection{Total transmitted/reflected flux}

Similarly, the integrated transmission or reflection at position $r$, $F(r)$ can be expressed in terms of polylogarithms, as (Eq. \ref{eq:RT} in main text):
\begin{eqnarray}
F(r)&=& \frac{-i \pi r}{2 \pi^2 r_0} \left(-\text{Li}_1 \left[ e^{-i \frac{\pi}{L}(r-r_0)} \right] + \text{Li}_1 \left[ e^{i \frac{\pi}{L}(r-r_0)} \right] + \text{Li}_1 \left[ e^{-i \frac{\pi}{L}(r+r_0-2r_{in})} \right] - \text{Li}_1 \left[ e^{i \frac{\pi}{L}(r+r_0-2r_{in})} \right] \right) \nonumber \\ 
&+& \frac{ L }{2 \pi^2 r_0} \left( \text{Li}_2 \left[ e^{-i \frac{\pi}{L}(r-r_0)} \right] + \text{Li}_2 \left[ e^{i \frac{\pi}{L}(r-r_0)} \right] - \text{Li}_2 \left[ e^{-i \frac{\pi}{L}(r+r_0-2r_{in})} \right] - \text{Li}_2 \left[ e^{i \frac{\pi}{L}(r+r_0-2r_{in})} \right] \right)
\end{eqnarray}
which gives, in the practical example, $R=|F(r=r_{in,p})|=0.95141$ and $T=|F(r=r_{in,p}+L_p)|=0.04859$.

\subsection{Comparison with Monte Carlo simulations}
The validity of our theoretical findings have been checked via Monte Carlo simulations of random walks in spherical shells. In these simulations, $10^6$ random walkers are launched one mean free path away from the inner boundary of the spherical shell (inside the spherical shell). Thereafter random walkers take isotropic, independent and exponentially distributed steps with average length $\ell$. Letting the numerical example from above serve as an example, we set $\ell$ to 1 $\upmu$m and the walker velocity to $v=200~\upmu$m/ps (resulting in a diffusion constant of $D=v\ell/3=67~\upmu$m$^2$/ps). The MC estimates of reflection and transmission and their respective characteristic times are shown in Table \ref{tab:MC_vs_theory} below along with the theoretical values presented above.
\begin{table}[htdp]
\caption{Comparison between the theoretical predictions and the outcome of random walk Monte Carlo. The simulated system is defined by $r_{in,p}=50~\upmu$m, $L_p=100~\upmu$m, $\ell=1\upmu$m and $D=67~\upmu$m$^2$/ps. Boundary reflections are set to zero (index-matched conditions). The source is placed at $r_0=r_{in,p}+\ell=51~\upmu$m.). Simulated values (MC estimates) are reported by stating the mean and standard deviation of three values each being the result of a simulation of $10^6$ random walkers.}
\begin{center}
  \begin{tabular}{lll}
    \hline
    Parameter & Theoretical value & MC estimate \\ 
    \hline
    $R$ (\%)          & 95.14    & 95.04$\pm$0.02 \\ 
    $\tau_R$ (ps) & 0.834    & 0.853$\pm$0.002 \\ 
    $T$ (\%)          & 4.86       & 4.96$\pm$0.02 \\ 
    $\tau_T$ (ps)  & 25.661 & 25.78 $\pm$0.07 \\  
    \hline
  \end{tabular}
\end{center}
\label{tab:MC_vs_theory}
\end{table}%

Clearly, the theoretical values are in very good agreement with simulation outcome. Also the shapes of the time-resolved reflection and transmission agree well, as shown in Figure \ref{fig:FIG_APP_Theory_vs_MC} below. Of course, more elaborate investigations of the validity of the diffusion model for, e.g., non-zero boundary conditions (mismatch in refractive index) and thin shells \--- extensively studied for common geometries such as slabs or semi-infinite media \--- remains to be done.
\begin{figure}[h]
\begin{center}
	\includegraphics[width=0.55\textwidth]{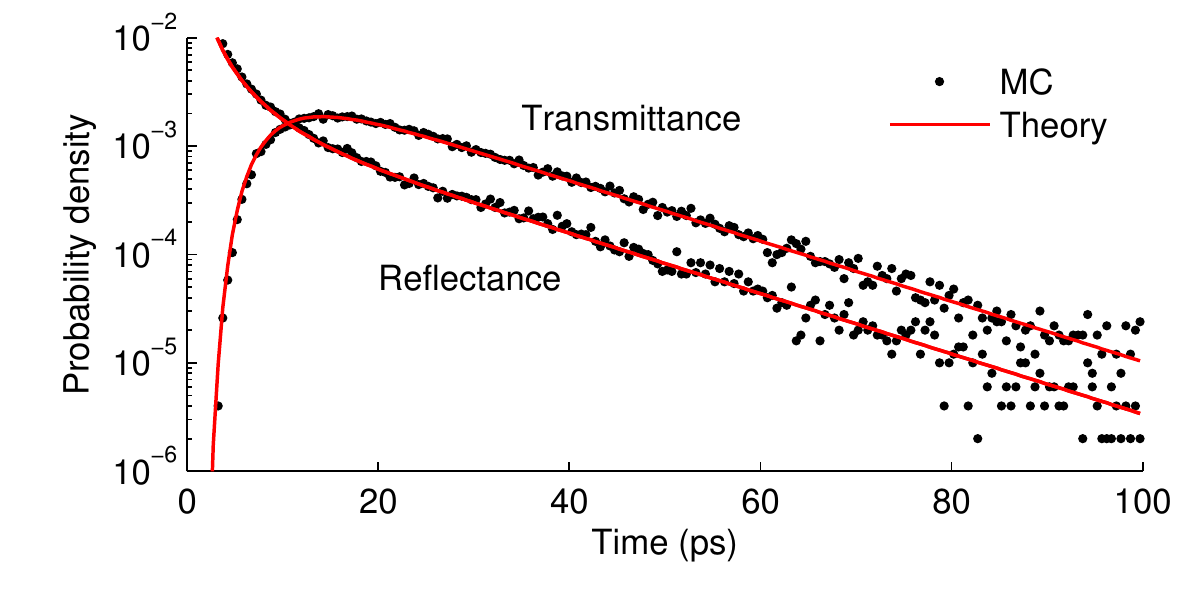}
\end{center}	
\caption{A comparison, in terms of time-resolved transmittance and reflectance, between the presented theory of diffusion in spherical shell and direct Monte Carlo simulation. The simulation data shown originates from one of the three sets of $10^6$ random walkers behind Table \ref{tab:MC_vs_theory}. For this set, reflectance was $R=95.06$\%, $\tau_R=0.852$ ps, transmittance $T=4.94$\%, and $\tau_T=25.75$ ps.
\label{fig:FIG_APP_Theory_vs_MC}}
\end{figure}

\newpage

\section{II. Ceramics Manufacturing}

An aqueous suspension with a solids loading of 50vol\% of ZrO$_2$ (TZ3YSE, Tosoh, Japan) and 0.3wt\% of dispersant (Dolapix PC 75, Zschimmer-Schwarz, Germany) was prepared by ball milling with milling media of zirconia. The suspension was diluted with water to a solids loading of 25vol\% and 6vol\% of latex as binder was introduced in form of a latex emulsion (LDM 7651S, Celanese, Sweden) with a particle size of 150 nm. For the holey ceramic, polyethylene microspheres (Cospheric, USA) with a size of 180$\upmu$m were added to the suspension. To avoid segregation of the polyethylene microspheres in the aqueous suspension, xanthan gum (Rhodophol 23, Rhodia) was used as a thickener to increase the viscosity of the suspension. For the holey ceramic, the volume of polyethylene microspheres used corresponded to 45 vol\% with respect to the total solid volume of zirconia and polyethylene microspheres. To maintain the homogeneity of the suspension with zirconia particles, latex emulsion and polyethylene micro spheres, the suspension was frozen drop by drop in liquid nitrogen followed by a freeze drying procedure to remove the ice by sublimation. The freeze dried particles were used to prepare ceramic green bodies by compaction. When the green bodies were sintered at 900$^\circ$C for two hours in a SiC furnace (Entech, Sweden), the organic additives were removed. Furthermore, the temperature was sufficient for an initial solid state diffusion, which allowed neck formation between the zirconia particles. At the same time, the temperature was not high enough to cause any sintering shrinkage of the powder compact. In this manner, a nanoporous ceramic with embedded macropores was obtained. As a reference, a material without any microsphere were also manufactured (same manufacturing procedure). 

Density measurements, performed with ArchimedesÕ method, showed that the nanoporous reference has a porosity of around 46\%. The holey ceramic, on the other hand, has a porosity of about 63.5\%. This is in good agreement with what is expected from the added fraction of microspheres. The 45vol\% microspheres should, when the other 55vol\% solids has formed a nanoporous media with 46\% porosity, give rise to a macroporosity of around 30\% and an overall porosity (macropores and nanoporosity) of about 63\%.

\section{III. Optical time-of-flight experiments}

\begin{figure}[h]
\begin{center}
	\includegraphics[width=0.65\textwidth]{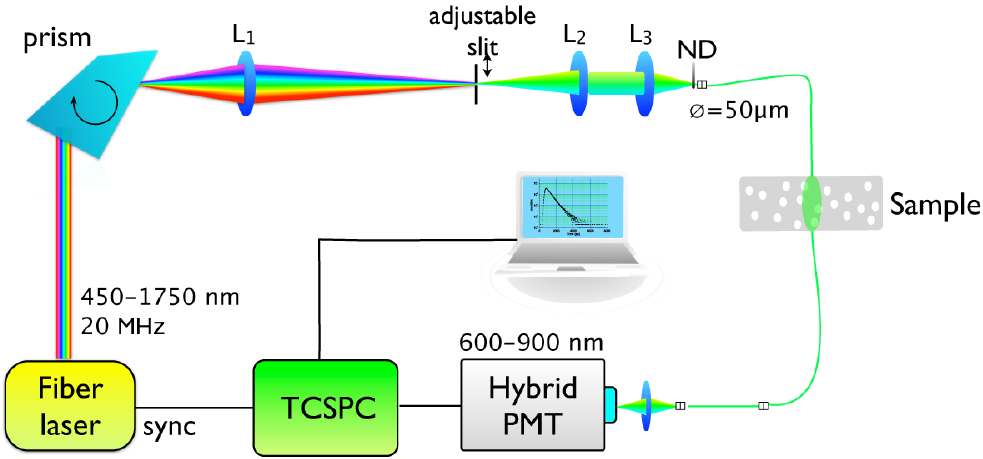}
\end{center}	
\caption{Schematic of the system used for optical time-of-flight spectroscopy (cf. \cite{Bassi2007_OptExpress,Bargigia2012_ApplSpectrosc}).
\label{fig:FIG_APP_TOF_System}}
\end{figure}

The system used for experiments is depicted in Fig.~\ref{fig:FIG_APP_TOF_System} and has been described in detail in \cite{Bassi2007_OptExpress,Bargigia2012_ApplSpectrosc}. It consists of a supercontinuum source (SuperK Extreme, NKT) emitting mode-locked laser pulses in the range 450-1750 nm at a repetition rate of 20 MHz. The white light exiting the source is dispersed by an SF10 Pellin-Broca prism and then focused on a variable slit by a 150 mm focal length achromatic doublet for spectral bandwidth selection. Tuning is achieved by the rotation of the prism. The slit plane is imaged on a 50 $\upmu$m graded index fiber by means of two achromatic lenses. The spectral bandwidth of the system ranges from about 3 nm at 600 nm to 6 nm at 900 nm. Light is delivered to and collected from the sample by means of 1 mm step-index fibers. The detector consists of a Hybrid PMT (HPM-100-50, Becker and Hickl, Germany). The instrumental response function (IRF), measured by setting the detection and injection fibers face-to-face, has a full-width half-maximum (FWHM) of about 180 ps over the whole spectral range. The time-of-flight (TOF) distribution of detected photons is measured by a time-correlated single photon counting (TCSPC) board (SPC-130, Becker and Hickl, Germany) mounted on the PC, which controls both the prism rotation and the data acquisition. 

The porous ceramics was measured in transmittance geometry with co-linear fiber optics. TOF distributions was acquired in 600-900 nm range 600-900 nm (in steps of 20 nm) at an intensity that gave about 400000 counts/s. The samples were around 3 mm thick, resulting in average TOF of a few nanoseconds, depending on the wavelength (e.g. TOF curves are far wider than the IRF). The size of the optical fiber collecting transmitted light was taken into account during evaluation of TOF distribution (i.e. during the assessment of the diffusion constant).
\end{document}